\documentclass[pre,twocolumn,
superscriptaddress,
floatfix]{revtex4-1}

\usepackage{amsmath}

\usepackage[english]{babel}

\usepackage{graphicx}
\usepackage[abs]{overpic}

\newcommand{\fix}[1]			
             {#1}

\begin{document}


\title{Do the rich get richer? An empirical analysis of the Bitcoin transaction network}
\author{D\'aniel Kondor}
\email{kdani88@elte.hu}
\affiliation{
Department of Physics of Complex Systems, E\"otv\"os Lor\'and University, Hungary\\
H-1117 Budapest, P\'azm\'any P\'eter S\'et\'any 1/A}
\author{M\'arton P\'osfai}
\affiliation{
Department of Physics of Complex Systems, E\"otv\"os Lor\'and University, Hungary\\
H-1117 Budapest, P\'azm\'any P\'eter S\'et\'any 1/A}
\affiliation{
Department of Theoretical Physics, Budapest University of Technology and Economics, \\
Budafoki \'ut 8, Budapest H--1111, Hungary}
\author{Istv\'an Csabai}
\affiliation{
Department of Physics of Complex Systems, E\"otv\"os Lor\'and University, Hungary\\
H-1117 Budapest, P\'azm\'any P\'eter S\'et\'any 1/A}
\author{G\'abor Vattay}
\affiliation{
Department of Physics of Complex Systems, E\"otv\"os Lor\'and University, Hungary\\
H-1117 Budapest, P\'azm\'any P\'eter S\'et\'any 1/A}

\date{\today}

\maketitle

\section*{Abstract}
	The possibility to analyze everyday monetary transactions is limited by the scarcity of available data, as this kind of
	information is usually considered highly sensitive. Present econophysics models are usually employed on presumed random networks of
	interacting agents, and only some macroscopic properties (e.g.~the resulting wealth distribution) are compared to real-world
	data. In this paper, we analyze Bitcoin, which is a novel digital currency system, where the complete list of transactions
	is publicly available. Using this dataset, we reconstruct the network of transactions and extract the time and amount of each payment.
	We analyze the structure of the transaction network by measuring network characteristics over time, such as the degree distribution, degree correlations and clustering. We find that linear preferential attachment drives the growth of the network. We also study the dynamics taking place on the transaction network, i.e. the flow of money. We measure temporal patterns and the wealth accumulation. Investigating the microscopic statistics of money movement, we find that sublinear preferential attachment governs the evolution of the wealth distribution. We report a scaling law between the degree and wealth associated to individual nodes.


\section{Introduction}

	In the past two decades, network science \fix{has successfully contributed to} many diverse scientific fields. Indeed, many complex systems can be 
	represented as networks, ranging from biochemical systems, through the Internet and the World Wide Web, to various social
	systems~\cite{barabasi_review,newman_reveiw,barabasibio,pastor_book,vespignani,havlin,critnet}. Economics also made use of the concepts of
	network science, gaining additional insight to the more traditional
	approach~\cite{networkfinancial1,networkfinancial2,networkfinancial3,networkfinancial4,networkfinancial5,palla}.
	Although a large volume of financial data is available for research, information about the everyday transactions of individuals is
	usually considered sensitive and is kept private. In this paper, we analyze Bitcoin, a novel currency system, where the complete list
	of transactions is accessible. We believe that this is the first opportunity to investigate the movement of money in such detail.
	
	Bitcoin is a decentralized digital cash system, there is no single overseeing authority~\cite{satoshi}. The system operates as an online
	peer-to-peer network, anyone can join by installing a client application and connecting it to the network. The unit of the
	currency is one bitcoin (abbreviated as BTC), and the smallest transferable amount is $10^{-8} \, \textrm{BTC}$. Instead of having a
	bank account maintained by a central authority, each user has a Bitcoin address, that consists of a pair of public and private keys.
	Existing bitcoins are associated to the public key of their owner, and outgoing payments have to be signed by the owner using his
	private key. To maintain privacy, a single user may use multiple addresses. Each participating node stores the complete list of previous
	transactions. Every new payment is announced on the network, and the payment is validated by checking consistency with the entire
	transaction history. To avoid fraud, it is necessary that the participants agree on a single valid transaction history. This process is 
	designed to be computationally difficult, so an attacker can only hijack the system if he possesses the majority of the computational power \fix{of participating parties}.
	Therefore the system is more secure if more resources are devoted to the validation process. To provide incentive, new bitcoins are created 
	periodically and distributed among the nodes participating in these computations. Another way to obtain bitcoins is to purchase them from 
	someone who already has bitcoins using traditional currency; the price of bitcoins is completely determined by the market.
	
	The Bitcoin system was proposed in 2008 by Satoshi Nakamoto, and the system went online in January 2009~\cite{satoshi,nakamoto,shamir,anonimity,structure}.
	For over a year, it was only used by a few enthusiasts, and bitcoins did not have any real-world value. \fix{A trading website called  MtGox was started
	in 2010,} making the exchange of bitcoins and conventional money significantly easier. More people and services joined the system, resulting a
	steadily growing exchange rate. Starting from 2011, appearances in the mainstream media drew wider public attention, which led to skyrocketing
	prices accompanied by large fluctuations~(see Fig.~\ref{addr1}). Since the inception of Bitcoin over 17 million transactions took place, and
	currently the market value of all bitcoins in circulation exceeds 1 billion dollars.  See the Methods section for more details of the system
	and the data used in our analysis.

	We download the complete list of transactions, and reconstruct the transaction network: each node represents a Bitcoin address, and
	we draw a directed link between two nodes if there was at least one transaction between the corresponding addresses.
	In addition to the topology, we also obtain the time and amount of every payment. Therefore, we are able to analyze both the evolution of the
	network and the dynamical process taking place on it, i.e. the flow and accumulation of bitcoins. To characterize the underlying 
	network, we investigate the evolution of basic network characteristics over time, such as the degree distribution, degree correlations and
	clustering. Concerning the dynamics, we measure the wealth statistics and the temporal patterns of transactions. To explain the
	observed degree and wealth distribution, we measure the microscopic growth statistics of the system. We provide evidence that preferential
	attachment is an important factor shaping these distributions. Preferential attachment is often referred to as the ``rich get richer''
	scheme, meaning that hubs grow faster than low-degree nodes. In the case of Bitcoin, this is more than an analogy: we find that the wealth
	of \fix{already rich nodes} increases faster than the wealth of \fix{nodes} with low balance; furthermore, we find positive correlation
	between the wealth and the degree of a node.

\section{Results}

\subsection{Evolution of the transaction network}

\begin{figure}
	\includegraphics{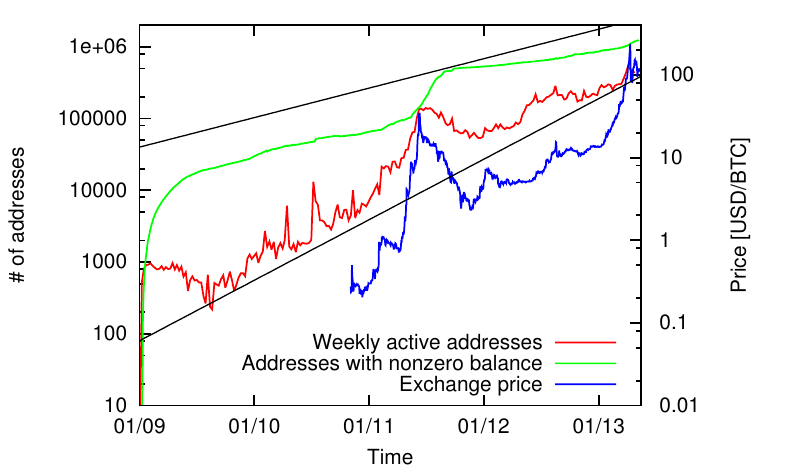}
	\caption{{\bf The growth of the Bitcoin network.} Number of addresses with nonzero balance (green), addresses in
		participating in at least one transaction in one week intervals (red) and the exchange price of bitcoins
		in US dollars according to MtGox, the largest Bitcoin exchange site (blue).  \fix{The black lines
		are exponential functions bounding the growth of the network size; the characteristic times are $188$ and
		$386$ days.}}
	\label{addr1}
\end{figure}

Bitcoin is an evolving network: new nodes are added by creating new Bitcoin addresses, and links are created if there is a transaction between 
two previously unconnected addresses. The number of nodes steadily grows over time with some fluctuations; especially noticeable is the
large peak which coincides with the first boom in the exchange rate in 2011 (Fig.~\ref{addr1})~\cite{structure}. After five years Bitcoin now has
$N=13,086,528$ nodes and $L=44,032,115$ links. To study the evolution of the network we measure the change of network characteristics in
function of time. We identify two distinct phases of growth: (i) The \emph{initial phase} lasted until the fall of 2010, in this 
period the system had low activity and was mostly used as an experiment. The network measures are characterized by large fluctuations.
(ii) After the initial phase the Bitcoin started to function as a real currency, bitcoins gained real value. The network measures
converged to their typical value by mid-2011 and they did not change significantly afterwards. We call this period the \emph{trading phase}.

\begin{figure}
	\includegraphics{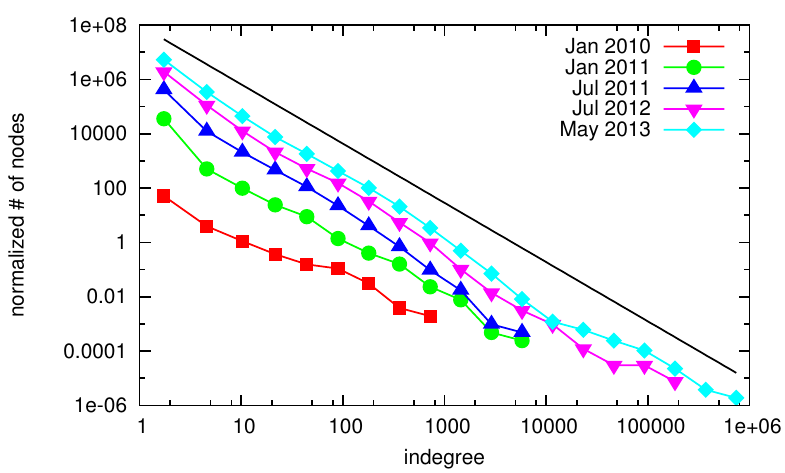}
	\caption{{\bf Evolution of the indegree distribution.} Since the beginning of 2011, the shape of the distribution
		does not change significantly. The black line shows a fitted power-law for the final network; the exponent
		is $2.18$. \fix{The data is log-binned for ease of visual inspection, the power-law is fitted on the original data~\cite{powerlaw}.}}
	\label{indegdist1}
\end{figure}

\begin{figure}
	\includegraphics{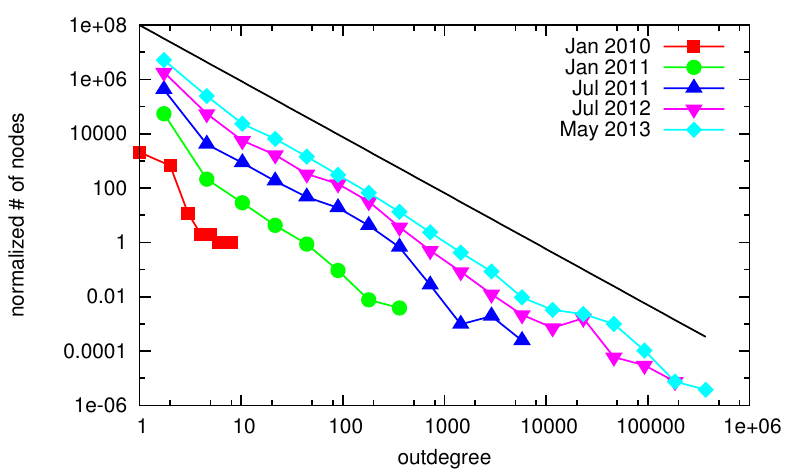}
	\caption{{\bf Evolution of the outdegree distribution.} The black line shows a fitted power-law for
		the final network; the exponent is $2.06$. \fix{The data is log-binned for ease of visual inspection, the power-law is fitted on the original data~\cite{powerlaw}.}}
	\label{outdegdist1}
\end{figure}

\begin{figure}
	\includegraphics{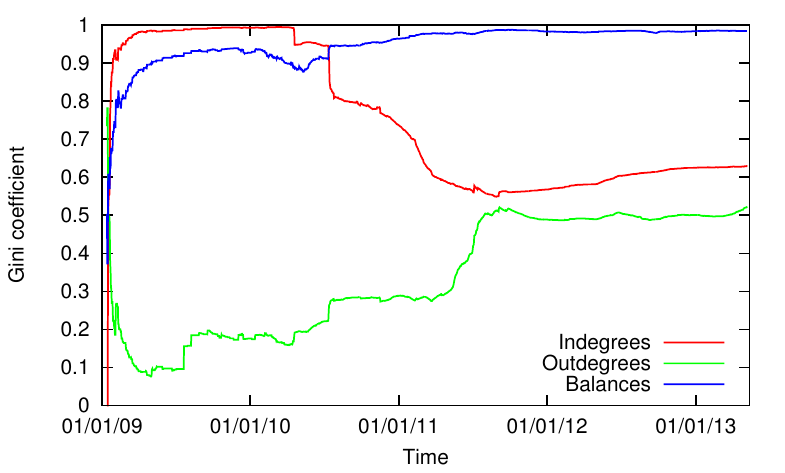}
	\caption{{\bf Evolution of the Gini coefficient of the degree and the balance distributions.} We observe the distinct initial
		phase lasting until mid-2011. The trading phase is characterized by approximately constant coefficients.}
	\label{ginitime}
\end{figure}

We first measure the degree distribution of the network. We find that both the in- and the outdegree distributions are highly heterogeneous, and they can be modeled with power-laws~\cite{powerlaw}. Figures~\ref{indegdist1} and~\ref{outdegdist1} show the distribution of indegrees and outdegrees at different points of time during the evolution of the Bitcoin network. In the initial phase the number of nodes is low, and thus fitting the data is prone to large error. In the trading phase, the exponents of the distributions do not change significantly, and they are approximated by power-laws $p_\text{in}(k_\text{in}) \sim k_\text{in}^{-2.18}$ and $p_\text{out}(k_\text{out}) \sim k_\text{out}^{-2.06}$.

To further characterize the evolution of the degree distributions we calculate the corresponding Gini coefficients in function of time (Fig.~\ref{ginitime}). The Gini coefficient is mainly used in economics to characterize the inequality present in the distribution of wealth, but it can be used to measure the heterogeneity of any empirical distribution. In general, the Gini coefficient is defined as
\begin{equation}
G = \frac{2 \sum_{i=1}^n i x_i}{n \sum_{i=1}^n x_i} -\frac{n+1}{n}
\end{equation}
where $\{x_i\}$ is a sample of size $n$, and $x_i$ are monotonically ordered, i.e.~$x_i \leq x_{i+1}$. $G=0$ indicates perfect equality,
\fix{i.e.~every node has the same wealth; and $G=1$ corresponds to complete inequality, i.e.~the complete wealth in the system is owned by a
single individual. For example, in the case of pure power-law distribution with $\alpha \geq 2$ exponent, the Gini coefficient is
$G = 1 / (2 \alpha - 3)$~\cite{paretogini}. This shows the fact that smaller $\alpha$ exponents yield more heterogeneous wealth distributions.}

In the Bitcoin network we find that in the initial phase the Gini coefficient of the indegree distribution is close to 1 and for the outdegree distribution it is much lower. We speculate that in this phase a few users collected bitcoins, and without the possibility to trade, they stored them on a single address. In the second phase the coefficients quickly converge to $G^\text{in}\approx 0.629$ and $G^\text{out}\approx 0.521$, indicating that normal trade is characterized by both highly heterogeneous in- and outdegree distributions.

\begin{figure}
	\includegraphics{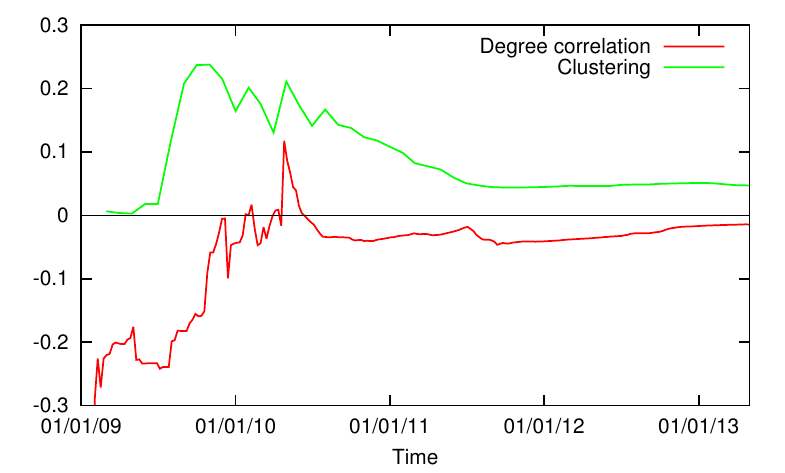}
	\caption{{\bf Evolution of the clustering coefficient and the out-in degree correlation coefficient.}
		\fix{After the initial phase, both measures reach a stationary value.}}
	\label{degcorrcltime}
\end{figure}

\begin{figure}
\centering
	\includegraphics{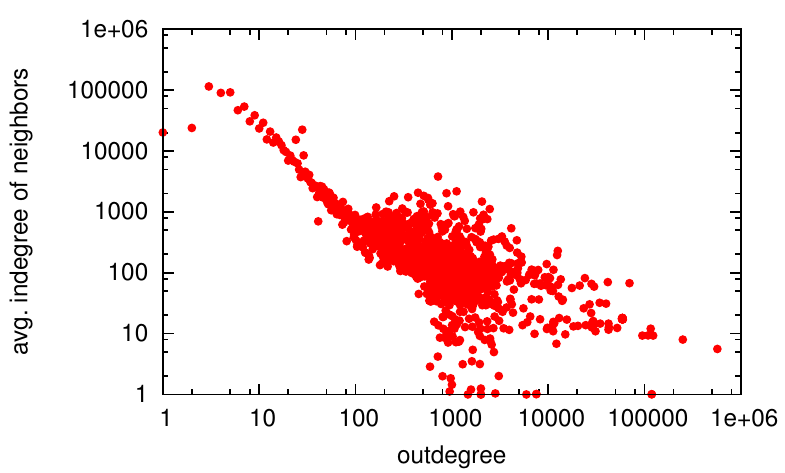}
	\caption{{\bf The average indegree of neighbors in the function of the outdegree $k^\text{in}_\text{nn}(k^\text{out})$}. \fix{In networks without degree correlations, the degree of connected nodes do not depend on each other, therefore for such networks we expect that $k^\text{in}_\text{nn}(k^\text{out})$ is constant. In the case of the Bitcoin network, we observe a clear disassortative
		behavior: $k^\text{in}_\text{nn}(k^\text{out})$ is a decreasing function, indicating that nodes with high outdegree tend to connect to nodes with low indegree.}}
	\label{anndoutin}
\end{figure}

To characterize the degree correlations we measure the Pearson correlation coefficient of the out- and indegrees of connected node pairs:
\begin{equation}
 r = \frac{\sum_{e} ( j^{\textrm{out}}_e-\overline{j^{\textrm{out}}} )
		( k^{\textrm{in}}_e-\overline{k^{\textrm{in}}} )}
	{ L \sigma_\text{out}\sigma_\text{in} }.
\end{equation}
Here $j^{\textrm{out}}_i$ is the outdegree of the node at the \emph{beginning} of link $e$, and $k^{\textrm{in}}_i$ is the indegree of the node at the
\emph{end} of link $e$. The summation $\sum_{e }\cdot$ runs over all links, $\overline{k^{\textrm{in}}}= \sum_{e} k^{\textrm{in}}_e / L$ and
$\sigma_\text{in}^2 = \sum_{e} ( k^{\textrm{in}}_e-\overline{k^{\textrm{in}}} )^2 / L$. We calculate $\sigma_\text{out}$ and
$\overline{j^{\textrm{out}}}$ similarly.

We find that the correlation coefficient is negative, except for only a brief period in the initial phase. After mid-2010, the degree correlation
coefficient stays between $-0.01$ and $-0.05$, reaching a value of $r\approx-0.014$ by 2013, suggesting that the network is disassortative
(Fig.~\ref{degcorrcltime}).
However, small values of $r$ are hard to interpret: it was shown that for large purely scale-free networks $r$ vanishes as the network size
increases~\cite{menche}. Therefore we compute the average nearest neighbor degree function $k^\text{in}_\text{nn}(k^\text{out})$ for the final network;
$k^\text{in}_\text{nn}(k^\text{out})$ measures the average indegree of the neighbors of nodes with outdegree $k^\text{out}$. We find clear disassortative
behavior (Fig.~\ref{anndoutin}).

We also measure the average clustering coefficient
\begin{equation}
 C = \frac1N \sum_v \frac{\Delta_v}{d_v (d_v-1)/2},
\end{equation}
which measures the density of triangles in the network. Here the sum $\sum_v\cdot$ runs over all nodes, and $\Delta_v$ is the
number of triangles containing node $v$. To calculate $\Delta_v$ we ignored the directionality of the links; $d_v$ is the
degree of node $v$ in the undirected network.

\begin{figure}
	\centering
	\begin{overpic}[unit=1mm]{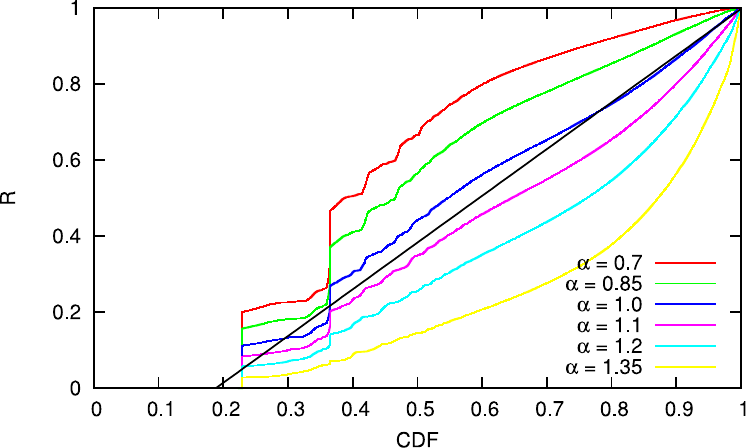}
		\put(12,27){\includegraphics{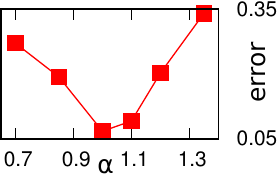}}
	\end{overpic}
	\caption{{\bf Rank function for new \fix{link creation}.} The cumulative distribution function of the $R$ values (see Eq.~\ref{erank})
		for exponents $0.7, 0.85, 1, 1.1, 1.2$ and $1.35$. The inset shows the \fix{Kolmogorov-Smirnoff} error for these exponents.}
	\label{pa6a10y0n}
\end{figure}

In the initial phase $C$ is high, fluctuating around $0.15$ (see Fig.~\ref{degcorrcltime}),
possibly a result of transactions taking place between addresses belonging to a few enthusiasts trying out the Bitcoin
system by moving money between their \fix{own} addresses. In the trading phase, the clustering coefficient reaches
a stationary value around $C\approx 0.05$, which is still higher than the clustering coefficient for random networks with
the same degree sequence ($C_\text{rand} \approx 0.0037(9)$).

	To explain the observed broad degree distribution, we turn to the microscopic statistics of link formation. Most real complex networks
	exhibit distributions that can be approximated by power-laws. Preferential attachment was introduced as a possible mechanism 
	to explain the prevalence of this property~\cite{barabasinew}. Indeed, direct measurements confirmed that preferential attachment
	governs the evolution of many real systems, e.g. scientific citation networks~\cite{newman,barabasipa,perc2}, collaboration
	networks~\cite{barabasipa2}, social networks~\cite{paonline,youtube} \fix{or language use~\cite{perc1}}. In its original form,
	preferential attachment describes the process when the probability 
	of forming a new link is proportional to the degree of the target node~\cite{barabasi}. In the past decade, several generalizations and 
	modifications of the original model were proposed, aiming to reproduce further structural characteristics of real systems~\cite
	{krapivsky, dorogovstev1, dorogovstev2, vazquez2}. Here, we investigate the nonlinear preferential attachment model~\cite{krapivsky},
	where the probability that a new link connects to node $v$ is
\begin{equation}
\label{eq:nonlinkernel}
 \pi(k_v) = \frac{k_v^\alpha}{\sum_w k_w^\alpha},
\end{equation}
	where $k_v$ is the indegree of node $v$, and $\alpha > 0$. The probability that the new link 
	connects to \emph{any} node with degree $k$ is $\Pi(k)\sim n_k (t) \pi(k)$, where $n_k (t)$ is the number of nodes with $k$ degree at
	the time of the link formation. We cannot test directly our assumption, because $\Pi(k)$ changes over time. To proceed we transform
	$\Pi(k)$ to a uniform distribution by calculating the rank function $R(k,t)$ for each new link given $\pi(k)$ and $n_k(t)$:
\begin{equation}
	R(k,t) = \frac{ \sum_{j=0}^k n_{j} (t) j^{\alpha} }{ \sum_{j=0}^{k_{\mathrm{max}}} n_{j} (t) j^{\alpha} } = %
	\frac{ \sum_{k_v < k} k_v^{\alpha} }{ \sum_v k_v^{\alpha} } \textrm{.}
	\label{erank}
\end{equation}

	If Eq.~\ref{eq:nonlinkernel} holds, $R(k,t)$ is uniformly distributed in the interval $[0,1]$, independently of $t$.
	Therefore, if we plot the cumulative distribution function, we get a straight line for the correct exponent $\alpha$. To
	determine the best exponent, we compare the empirical distribution of the $R$ values to the uniform distribution for
	different exponents by computing the Kolmogorov-Smirnoff distance between the two distributions.
	
	Evaluating our method for indegree distribution of the Bitcoin network, we find good correspondence between the empirical
	data and the presumed conditional probability function; the exponent giving the best fit is $\alpha \approx 1$
	(Fig.~\ref{pa6a10y0n}). This shows that the overall growth statistics agree well with the preferential attachment
	process. Of course, preferential attachment itself cannot explain the disassortative degree correlations and the high
	clustering observed in the network. We argue that preferential attachment is a key factor shaping the degree distribution,
	however, more detailed investigation of the growth process is necessary to explain the higher order correlations.

\subsection{Dynamics of transactions}

\begin{figure}
\centering
	\includegraphics{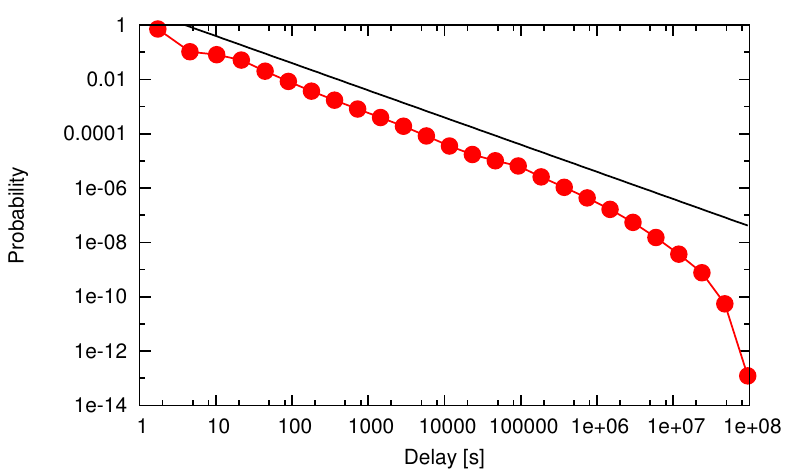}
	\caption{{\bf Distribution of time delay between transactions initiated from a single Bitcoin address.}
		We observe a power-law distribution close to the widely observed $P(T) \sim T^{-1}$, the exponential cutoff
		corresponds to the finite lifetime of the Bitcoin system.
		}
	\label{dtin2}
\end{figure}

\begin{figure}
\centering
	\includegraphics{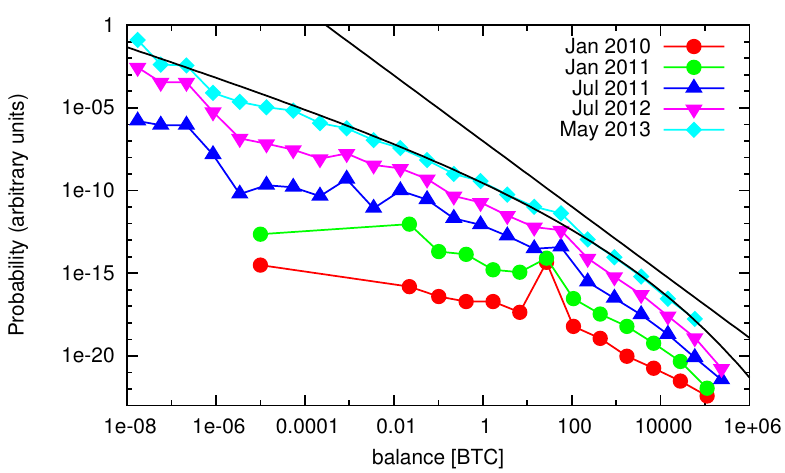}
	\caption{{\bf Evolution of the distribution of balances of individual Bitcoin addresses.} \fix{The distributions are shifted
		by arbitrary factors along the vertical axis for better visibility of the separate lines. The black lines are stretched exponential
		and power-law fits of the last empirical distribution. The tail can be approximated by a power-law with exponent $-1.984$, however, the rest of the fit is unsatisfactory. Therefore, we fit the distribution with a stretched exponential distribution of form 
		$P(b) \sim b^{-\gamma} e^{-(a b)^{1-\gamma}}$. We find a better approximation of the whole distributions; the parameters are $\gamma = 0.873$ and $a = 8014 \, \textrm{BTC}^{-1}$.
		}}
	\label{balanceslb}
\end{figure}

In the this section, we analyze the detailed dynamics of money flow on the transaction network. The increasing availability of digital traces
of human behavior revealed that various human activities, e.g. mobility patterns, phone calls or email communication, are often characterized
by heterogeneity~\cite{bursts2, kertesz, mobility, mobile}. Here we show that the handling of money is not
an exception: we find heterogeneity in both balance distribution and temporal patterns. We also investigate the microscopic
statistics of transactions.

The state of node $v$ at time $t$ is given by the balance of the corresponding address $b_v(t)$, i.e. the number of bitcoins associated to node $v$.
The transactions are directly available, and we can infer the balance of each node based on the transaction list. Note that the
overall quantity of bitcoins increases over time: Bitcoin rewards users devoting computational power to sustain the system.

We first investigate the temporal patterns of the system by measuring the distribution of inactivity times $T$. The inactivity time is defined as the time elapsed between two consecutive outgoing transactions from a node. We find a broad distribution that can be approximated by the power-law $P(T) \sim 1 / T$ (Fig.~\ref{dtin2}), in agreement with the behavior widely observed in various complex systems~\cite{bursts1, bursts2, poissonian_expl, temporal_review}.

It is well known that the wealth distribution of society is heterogeneous; the often cited --and quantitatively not precise-- 80-20 rule
of Pareto states that the top 20\% of the population controls 80\% of the total wealth. In line with this, we find that the wealth
distribution in the Bitcoin system is also highly heterogeneous. \fix{The proper Pareto-like statement for the Bitcoin system would be
that the 6.28\% of the addresses posesses the 93.72\% of the total wealth.} We measure the distribution of balances at different points
of time, and we find a stable distribution. The tail of wealth distribution is generally modeled with a power-law~\cite{boltzman,wealthchina,forbes},
following this practice we find a power-law tail $\sim x^{-1.984}$ for balances $\gtrsim 50 \textrm{BTC}$ (see Fig.~\ref{balanceslb}).
However, visual inspection of the fit is not convincing: the scaling regime spans only the last few orders of magnitude, and fails to
reproduce the majority of the distribution. Instead we find that the overall behavior is much better approximated by the stretched
exponential distribution $P(b) \sim b^{-\gamma} e^{-(a b)^{1-\gamma}}$, where $\gamma = 0.873$ and $a = 8014 \, \textrm{BTC}^{-1}$.

To further investigate the evolution of the wealth distribution we measure the Gini coefficient over time. We find that the distribution is characterized by high values throughout the whole lifetime of the network, reaching a stationary value around $G\approx 0.985$ in the trading phase (see Fig.~\ref{ginitime}).

To understand the origin of this heterogeneity, we turn to the microscopic statistics of acquiring bitcoins. Similarly to the case of degree distributions, the observed heterogeneous wealth distributions are often explained by preferential attachment. Moreover,
preferential attachment was proposed significantly earlier in the context of wealth distributions than
complex networks~\cite{simon}. In economics preferential attachment is traditionally called the Matthew effect or the
``rich get richer phenomenon''~\cite{matthew}. It states that the growth of the wealth of each individual
is proportional to the wealth of that individual. In line with this principle, several statistical
models were proposed to account for the heterogeneous wealth distribution~\cite{boltzman,wealthpow,pamoney1,pamoney2}.

\begin{figure}[t]
	\centering
	\includegraphics{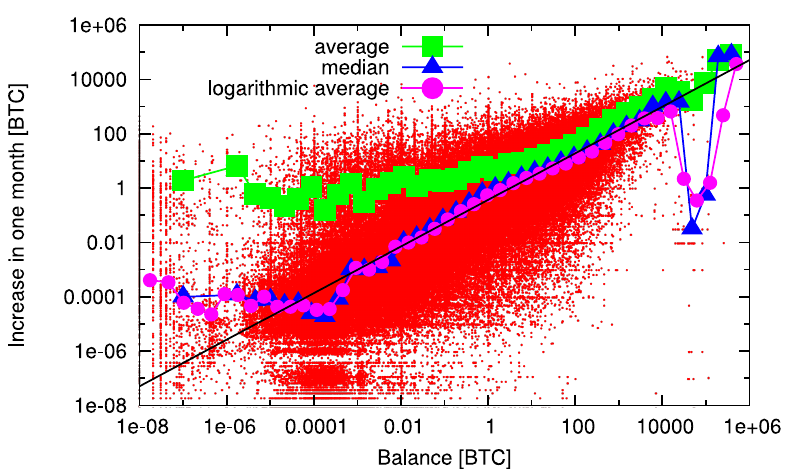} \\
	\includegraphics{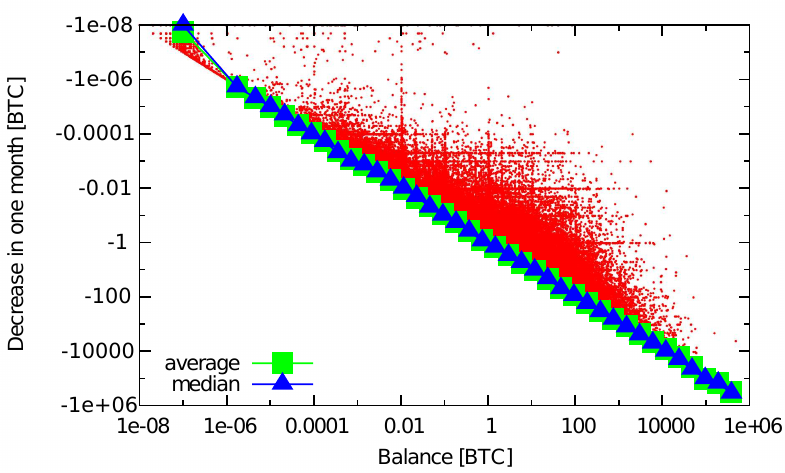}
	\caption{{\bf Change of balances in one month windows.} Increase (top) and decrease (bottom, vertical axis is inverted) of node balances
		in one month windows as a function of their balance at the beginning of each month.
		\fix{We show the raw data (red), the average (green), median (blue) and logarithmic average (magenta).
		The later three are calculated for logarithmically sized bins. We find a clear positive correlation:
		addresses with high balance typically increase their wealth more than addresses with low balance.
		The median and the logarithmic average values almost coincide, which suggests multiplicative
		fluctuations. The median and the logarithmic average increase approximately as power-laws for several orders
		of magnitude. The black line is a power-law fit for the double logarithmic data; the exponent is
		$0.857$.}}
	\label{pamoney}
\end{figure}

\fix{To find evidence supporting this hypothesis,} we first investigate the change of balances in fixed time windows. We calculate the difference between the balance of each address at the end and 
at the start of each month. We plot the differences in function of the starting balances (Fig.~\ref{pamoney}). When the balance increases, we 
observe a positive correlation: the average growth increases in function of the starting balance, \fix{and it is approximated by the power-law
$\sim b^{0.857}$}. This indicates the ``rich get richer'' 
phenomenon is indeed present in the system. For decreasing balances, we find that a significant number of addresses lose all their wealth in the 
time frame of one month. This phenomenon is specific to Bitcoin: due to the privacy concerns of users, it is generally considered a good practice 
to move unspent bitcoins to a new address when carrying out a transaction~\cite{newaddress}. 

\begin{figure}
	\centering
	\begin{overpic}[unit=1mm
			]{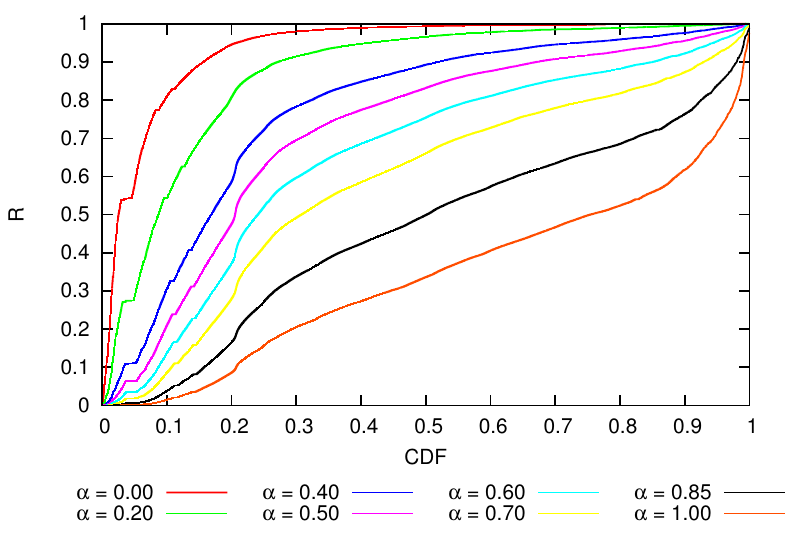}
		\put(46,14){\includegraphics{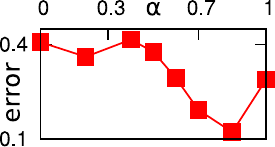}}
	\end{overpic}
	\caption{{\bf Rank function for the growth of balances.} 
	The cumulative distribution function of the $R$ values (see Eq.~\ref{erank})
		for exponents $0$, $0.2$, $0.4$, $0.5$, $0.6$, $0.7$, $0.85$ and $1$. The inset shows the maximum Kolmogorov-Smirnoff error for these exponents. Here, the results are not as obvious as in the case of link creation (Fig.~\ref{pa6a10y0n}; a simple power-law form like in Eq.~\ref{eq:nonlinkernel} is not
		sufficient to accurately model the statistics of money flow. On the other hand, the ``average'' behavior
		shows a correlation between the balance and the increase of the balance: the uncorrelated assumption ($\alpha = 0$) clearly
		gives a much worse approximate than the exponents that presume preferential attachment ($\alpha > 0$).}
	\label{addrbalct}
\end{figure}

To better quantify the preferential attachment, we carry out a similar analysis to the previous section. 
However, there is a technical difference: in the case of the evolution of the transaction network, for each event the degree of a node increases by exactly one. In the case of the wealth distribution there is no such constraint. To overcome this difficulty we consider the increment of a node's balance by one unit as an event, e.g. if after a transaction $b_v$ increased by $\Delta b_v$, we consider it as $\Delta b_v$ separate and simultaneous events.
We only consider events when the balance associated to an address increases, i.e. the address receives a payment. 
We now calculate the rank function
$R(b,t)$ defined in Eq.~\ref{erank}, and plot the cumulative distribution function of the $R$ values observed throughout
 the whole time evolution of the Bitcoin network (Fig.~\ref{addrbalct}).
Visual inspection shows that no single exponent provides a satisfying result, meaning that $\pi(b_v)$ cannot be modeled by a simple power-law relationship like in Eq.~\ref{eq:nonlinkernel}. However, we do find that the ``average''
behavior is best approximated by exponents around $\alpha \approx 0.8$, suggesting that $\pi(b_v)$ is a sublinear function. In the context of network evolution, previous theoretical work found that sublinear preferential attachment leads to a stationary stretched exponential distribution~\cite{krapivsky}, in line with our observations.

\begin{figure}[t]
	\centering
	\begin{overpic}[unit=1mm
			]{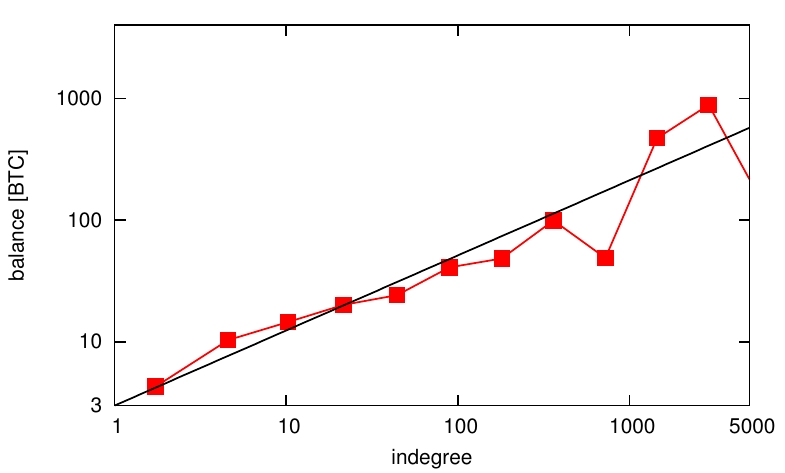}
		\put(9,23){\includegraphics{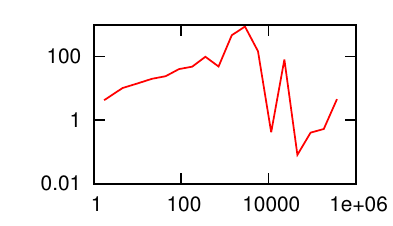}}
	\end{overpic}
	\caption{{\bf Average node balances as a function of the indegrees.} We calculate the averages for logarithmically sized bins. We find strong correlation between the balance and the indegree of individual nodes. The main plot shows indegree values up to $k_\text{in}\approx 3000$, only 75 nodes ($0.0063 \%$) have higher indegree, the averages calculated for such small sample result in high fluctuations (see inset).
		We also measure both the Pearson and Spearman correlation coefficient: The Pearson correlation coefficient of the full dataset is $0.00185041$, while the Spearman rank correlation
		coefficient is $0.275881$. \fix{(Note that the Pearson correlation coefficient
		measures the linear dependence between two variables, while the Spearman coefficient evaluates monotonicity)}.
		We test the statistical significance of the correlation by randomizing the dataset 1000 times and calculating the Spearman coefficient for each randomization. We find that the average Spearman coefficient is
		$10^{-4}$ with a standard deviation of $9.5 \cdot 10^{-4}$, indicating that the correlation is indeed significant.}
	\label{degbal}
\end{figure}

We have investigated the evolution of both the transaction network and the wealth distribution separately. However, it is clear
that the two processes are not independent. To study the connection between the two, we measure the correlation between the indegree 
and balance associated to the individual nodes. We plot the average balance of addresses as a function of their degrees on
Fig.~\ref{degbal}. For degrees in the range of $1$--$3000$ (over $99.99\%$ of all nodes with nonzero balance), the average balance
is a monotonously increasing function of the degree, and it is approximated by the power-law $b \sim k_\text{in}^{0.617}$, indicating
that the accumulated  wealth and the number of distinct transaction partners an individual has are inherently related. Similar scaling
was reported by Tseng et al., who conducted an online experiment where volunteers traded on a virtual market~\cite{pamoney2}.

\section{Methods}

\subsection{The Bitcoin network}
	
\begin{figure}
	\includegraphics{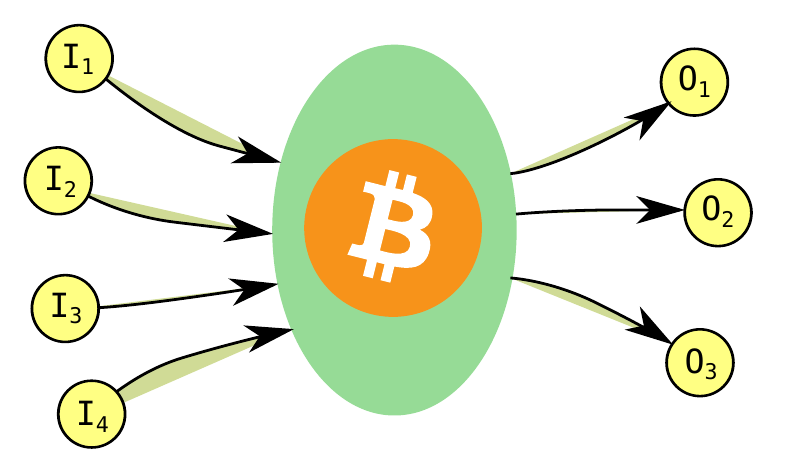}
	\caption{{\bf Schematic view of a Bitcoin transaction.} Here we have four input (\texttt{I$_1$}--\texttt{I$_4$}) and three
		output (\texttt{O$_1$}--\texttt{O$_3$}) addresses. \fix{Links in our analysis are created pointing from each input to each output address.}}
	\label{trasaction1}
\end{figure}

	Bitcoin is based on a peer-to-peer network of users connected through the Internet, where each node
	stores the list of previous transactions and validates new transactions based on a proof-of-work system. Users
	announce new transactions on this network, these transactions are formed into \emph{blocks} at an approximately constant rate
	of one block per 10 minutes; blocks contain a varying number of transactions. These blocks form the block-chain,
	where each block references the previous block. Changing a previous transaction (e.g. double spending) would require
	the recomputation of all blocks since then, which becomes practically infeasible after a few blocks. To send or
	receive bitcoins, each user needs at least one address, which is a pair of private and public keys. The public
	key can be used for receiving bitcoins (users can send money to each other referencing the recipient's public key),
	while sending bitcoins is achieved by signing the transaction with the private key. Each transaction consists of
	one or more \emph{inputs} and \emph{outputs}. In Fig.~\ref{trasaction1} we show a schematic view of a typical Bitcoin
	transaction. Readers interested in the technical details of the system can consult the original paper by Satoshi
	Nakamoto~\cite{satoshi} or the various resources available on the Internet~\cite{bcweb,clients}.

	An important aspect of Bitcoin is how new bitcoins are created, and how new users can
	acquire bitcoins. New bitcoins are generated when a new block is formed as a reward to the users
	participating in block generation. The generation of a valid new block involves solving a reverse hash
	problem, whose difficulty can be set in a wide range. Participating in block generation is referred to as
	\emph{mining} bitcoins. The nodes in the network regulate the block generation process by adjusting the
	difficulty to match the processing power currently available. As interest in the Bitcoin system grew, the
	effort required to generate new blocks, and thus receive the newly available bitcoins, has increased
	over 10 million fold; most miners today use specialized hardware, requiring significant investments.
	Consequently, an average Bitcoin user typically acquires bitcoins by either buying them at an exchange
	site or receiving them as compensation for goods or services.
	
	Due to the nature of the system, the record of all previous transactions since its beginning are publicly
	available to anyone participating in the Bitcoin network. From these records, one can recover the sending and receiving
	addresses, the sum involved and the approximate time of the transaction. Such detailed
	information is rarely available in financial systems, making the Bitcoin network a valuable source of empirical
	data involving monetary transactions. Of course, there are shortcomings: only the
	addresses involved in the transactions are revealed, not the users themselves. While providing complete
	anonymity is not among the stated goals of the Bitcoin project~\cite{anonimity2}, identifying addresses
	belonging to the same user can be difficult~\cite{anonimity}, especially on a large scale. Each
	user can have an unlimited number of Bitcoin addresses, which appear as separate nodes in the transaction records.
	When constructing the network of users, these addresses would need to be joined to a single entity.
	
	Another issue arises not only for Bitcoin, but for most online social datasets: It is hard to determine which observed phenomena are specific to the system, and which results are general. We do not know to what extent the group of people using the system can be considered as a representative sample of the society. In the case of Bitcoin for example, due to the perceived anonymity of the system, it is widely used for commerce of illegal items and substances~\cite{silkroad}; these types of transactions are probably overrepresented among Bitcoin transactions. Ultimately, the validity of our results will be tested if data becomes available from other sources, and comparison becomes possible.

\subsection{Data}
	We installed the open-source \texttt{bitcoind} client and downloaded the blockchain from the peer-to-peer
	network on May 7th, 2013. We modified the client to extract the list of all transactions in a human-readable format. We
	downloaded more precise timestamps of transactions from the \texttt{blockchain.info} website's archive. The data and the source code of the modified
	client program is available at the project's website~\cite{web} or through the Casjobs web database
	interface~\cite{nmvo,matray}.
		
	The data includes 235,000 blocks, which contain a total of 17,354,797 transactions. This dataset includes
	13,086,528 addresses (i.e. addresses appearing in at least one transaction); of these, 1,616,317 addresses
	were active in the last month. The Bitcoin network itself does not store balances associated with addresses,
	these can be calculated from the sum of received and sent bitcoins for each address; \fix{preventing overspending
	is done by requiring that the input of a transaction corresponds to the output of a previous transaction}.
	Using this method, we found that approximately one million addresses had nonzero balance at the time
	of our analysis.
	
\section{Discussion}

We have preformed detailed analysis of Bitcoin, a novel digital currency system. A key difference from traditional currencies handled by banks is the open nature of the Bitcoin: each transactions is publicly announced, providing unprecedented opportunity to study monetary transactions of individuals. We have downloaded and compiled the complete list of transactions, and we have extracted the time and amount of each payment. We have studied the structure and evolution of the transaction network, and we have investigated the dynamics taking place on the network, i.e. the flow of bitcoins.

Measuring basic network characteristics in function of time, we have identified two distinct phases in the lifetime of the system: (i) When the system was new, no businesses accepted bitcoins as a form of payment, therefore Bitcoin was more of an experiment than a real currency. This initial phase is characterized by large fluctuations in network characteristics, heterogeneous indegree- and homogeneous outdegree distribution. (ii) Later Bitcoin received wider public attention, the increasing number of users attracted services, and the system started to function as a real currency. This trading phase is characterized by stable network measures, dissasortative degree correlations and power-law in- and outdegree distributions. We have measured the microscopic link formation statistics, finding that linear preferential attachment drives the growth of the network.

To study the accumulation of bitcoins we have measured the wealth distribution at different points in time. We have found that this distribution is highly heterogeneous through out the lifetime of the system, and it converges to a stable stretched exponential distribution in the trading phase. We have found that sublinear preferential attachment drives the accumulation of wealth. Investigating the correlation between the wealth distribution and network topology, we have identified a scaling relation between the degree and wealth associated to individual nodes, implying that the ability to attract new connections and to gain wealth is fundamentally related.

We believe that the data presented in this paper has great potential to be used for evaluating and refining econophysics models,
as not only the bulk properties, but also the microscopic statistics can be readily tested. To this end,
we make all the data used in this paper available online to the scientific community in easily accessible
formats~\cite{web,nmvo,matray}.

\section*{Acknowledgments}
	
	\fix{The authors thank Andr\'as Bodor and Philipp H\"ovel for many useful discussions and suggestions.}
	This work has been supported by the European Union under grant agreement
	No. FP7-ICT-255987-FOC-II Project.
	The authors thank the partial support of
	the European Union and the European Social Fund through project FuturICT.hu
	(grant no.: TAMOP-4.2.2.C-11/1/KONV-2012-0013),
	the OTKA 7779 and the NAP 2005/KCKHA005 grants.
	EITKIC\_12-1-2012-0001 project was partially supported by the Hungarian Government,
	managed by the National Development Agency, and financed by the Research and
	Technology Innovation Fund and the MAKOG Foundation.



\begin{thebibliography}{99}
\bibitem{matthew}
		``For whosoever hath, to him shall be given, and he shall have more abundance: but whosoever hath not,
		from him shall be taken away even that he hath.'' The Bible, Matthew 13:12, King James translation
\end{thebibliography}
\end{document}